\documentclass[10pt,journal]{IEEEtran}

\usepackage{lineno,hyperref,psfrag,float,color}
\usepackage{tikz}
\usepackage{circuitikz}
\usetikzlibrary{arrows.meta}
\usepackage{amssymb}

\usepackage{lineno,hyperref,psfrag,float,color}
\usepackage{amsmath,amssymb,euscript ,yfonts,psfrag,latexsym,dsfont,graphicx,bbm,color,amstext,wasysym,subfig,flushend,parskip}
\usepackage{times,mathptmx,epsfig,graphicx,color,balance}
\usepackage{color}
\usepackage{graphicx}
\usepackage{psfrag}
\usepackage{epstopdf}
\usepackage{amssymb}
\usepackage{amsmath}
\usepackage{amsfonts}

\modulolinenumbers[5]

\newcommand{\blue}{\color{black}}
\newcommand{\red}{\color{black}}

\newtheorem{definition}{Definition}
\newtheorem{example}{Example}

\begin{document}

\title{Principles of lossless adjustable one-ports}

\author{Tryphon T. Georgiou, Faryar Jabbari and Malcolm C. Smith}
%\author{Tryphon T. Georgiou, Faryar Jabbari and Malcolm C. Smith~\IEEEmembership{Fellow,~IEEE}}
%\ead{tryphon@uci.edu}
%\author[UCI]{Faryar Jabbari}
%\ead{fjabbari@uci.edu}
%\address[UCI]{Mechanical and Aerospace Engineering, University of California, Irvine, CA 92697, USA}

%\author[mymainaddress]{Malcolm C. Smith\corref{mycorrespondingauthor}}
%\ead{mcs@eng.cam.ac.uk}
%\address[mymainaddress]{Department of Engineering, University of Cambridge, Cambridge CB2 1PZ, UK}

\markboth{lossless adjustable one-ports \hfill \today\hspace*{10pt}page\hspace*{3pt}}%
{Shell \MakeLowercase{\textit{et al.}}: Bare Demo of IEEEtran.cls for Journals}

\maketitle

\begin{abstract}
This paper explores the possibility to construct two-terminal mechanical devices (one-ports) which are lossless and adjustable.  To be lossless, the device must be passive (i.e.\  not requiring a power supply) and non-dissipative.  To be adjustable, a parameter of the device should be freely variable in real time as a control input.  
For the simplest lossless one ports, the spring and inerter, the question is whether the stiffness and inertance may be varied freely in a lossless manner.  We will show that the typical laws which have been proposed for adjustable springs and inerters are necessarily active and that it is not straightforward to modify them to achieve losslessness, or indeed passivity.  By means of a physical construction using a lever with moveable fulcrum we will derive device laws for adjustable springs and inerters which satisfy a formal definition of losslessness.  We further provide a construction method which does not require a power supply for physically realisable translational and rotary springs and inerters.  The analogous questions for lossless adjustable electrical devices are examined.
\end{abstract}


\section{Introduction}

Is it possible to build a spring with a ``workless knob" which freely adjusts its stiffness in real time?  Such a contrivance would behave like a conventional linear spring when the knob is stationary.  Energy imparted through compression or extension would be available for extraction again.  Adjustment of the knob would not involve any energy transfer between the environment and the contrivance.  Current methods to adjust the stiffness of springs do not answer this question, since they require active actuation, dissipation, or restrictive conditions on the switching of the spring constant.  We will provide an answer to this and related questions in the present paper.

The question is motivated by the ubiquity of the adjustable damper.  Such devices allow their proportionality constant to be adjusted, typically by a variable orifice controlled by a solenoid valve, or a magnetorheological fluid whose viscosity is altered by a magnetic field.  Adjustable dampers are much used for the control of mechanical systems, e.g.\ automotive suspensions \cite{butsuen1989optimal}, \cite{savaresi2010semi}, \cite{brezas2015clipped}, \cite{smith2018mclaren}.  The variable damper constant plays the role of a control input which may be adjusted by a control law that minimises a performance criterion.  Such devices are sometimes termed ``semi-active'' since a (small) power source is employed to effect the adjustment.  Nevertheless, the instantaneous power absorbed by the device can never be negative, and so from a terminal point of view it appears passive.  It is reasonable to expect that adjustable springs with similar properties would also offer performance advantages in a control system which would make them attractive in applications.  

An analogous question arises for the inerter \cite{smith2002synthesis}, which is a two-terminal mechanical device such that the equal and opposite force at the terminals is proportional to the relative acceleration between them.  The constant of proportionality is termed the inertance.  The question is whether an adjustable inerter is physically realisable as a lossless device, i.e.\ whether an inerter can be manufactured with a ``workless knob" which freely adjusts its inertance in real time.
%\red deleted"?" 

In the robotics field ``Variable Stiffness Actuators'' have been considered extensively (see \cite{vanderborght2013variable},  \cite{wolf2016variable} for recent surveys and the references therein).  {\red As noted in  
\cite{vanderborght2013variable} there are three principal methods to construct variable stiffness devices: adjustable spring preload; variable transmission or gearing ratio, including adjustments by a moveable pivot \cite{jafari2011awas,groothuis2014variable,liu2018modeling,lu2018theoretical}; change of physical properties of the spring.   
Each of the methods described requires some form of active force input, most commonly via electromechanical actuation.
%In the context of the present paper we highlight in particular
%work of the second type where the gearing ratio is adjusted by a moveable pivot  \cite{jafari2011awas,groothuis2014variable,liu2018modeling,lu2018theoretical}.
}

In \cite{bobrow1995active}, \cite{jabbari2002vibration} a passive ``resettable'' spring is proposed which requires minimal energy for switching.  A piston and cylinder arrangement acts in parallel with a conventional spring so that the closing of a valve allows the fluid in the cylinder to play the role of an additional spring.  In its simplest form this allows switching between two different levels of stiffness.  The closing of the valve (to increase the stiffness) can be effected at any time with minimal energy requirement.  The opening of the valve (to reduce the stiffness) is constrained to times at which there is no stored energy in the fluid, otherwise there is energy dissipation.  Control problems are considered which respect to the constraint on the timing of valve opening.

The possible benefits of adjustable inerters have been considered recently \cite{chen2014semi}, \cite{brzeski2015novel}, \cite{lazarek2018design}, \cite{garrido2018assumed}. 
A device law of a ``semi-active'' inerter is evaluated for a vehicle suspension system in \cite{chen2014semi} without considering the issue of realisability. 
In \cite{brzeski2015novel} a tuned mass damper (TMD) is proposed which incorporates an adjustable inerter making use of a rack and pinion and continuously variable transmission (CVT).  The CVT allows precise tuning of the natural frequency of the TMD, but energy requirements for the adjustment of the CVT are not considered.

In \cite{garrido2018assumed} the stability of control systems incorporating ``semi-active'' devices is considered.  It is pointed out that the commonly assumed device laws for ``variable-stiffness springs'' and ``variable-inertance inerters'' are in fact active, and that interconnections of such devices with passive elements may lead to instability.  A mechanical design for an (active) adjustable inerter is presented and studied in the context of vibration suppression of a building structure.  The potential benefits as well as the risk of instability are highlighted.

The work presented herein explores the existence of physically realizable device laws that are both lossless and adjustable, without essential restrictions on the values and timing of their control input parameter. Physical implementations of such device laws are envisioned as control components in applications areas that include the aforementioned areas of robotics, vibration suppression in buildings, and automotive suspension.  The control problems that result are expected to offer interesting technical challenges due to their non-linear character (e.g.\ as in \cite{brezas2015clipped} where the control input multiplies a state).  It is beyond the scope of this paper to explore these challenges and the potential performance benefits for specific applications.

The present paper is structured as follows. In Section~\ref{sec:mech} the basic definitions of mechanical one-ports, adjustability, passivity and losslessness are provided.  Section~\ref{sec:devicelaws} shows in a series of six examples that none of the commonly assumed device laws for adjustable springs or inerters or variants are lossless, and indeed all are non-passive, i.e.\ active.  Section~\ref{sec:planar} uses an idealised mechanical arrangement of a lever with moveable fulcrum to derive device laws for lossless adjustable springs and inerters.   Section~\ref{sec:canonical} presents a physical implementation of the moveable fulcrum concept without internal power source and introduces the names of varspring and varinerter for the canonical lossless adjustable spring and inerter.  Section~\ref{sec:rotary} presents a method for physical implementation of rotary varsprings and varinerters.  The paper concludes with a discussion of the analogous device laws in the electrical domain in Section~\ref{sec:elec}.

%\paragraph{Installation} If the document class \emph{elsarticle} is not available on your computer, you can download and install the system package \emph{texlive-publishers} (Linux) or install the \LaTeX\ package \emph{elsarticle} using the package manager of your \TeX\ installation, which is typically \TeX\ Live or Mik\TeX.
%
%\begin{itemize}
%\item document style
%\item baselineskip
%\item front matter
%\item keywords and MSC codes
%\item theorems, definitions and proofs
%\item lables of enumerations
%\item citation style and labeling.
%\end{itemize}

\section{Mechanical one-ports}\label{sec:mech}

\subsection{Definitions}\label{sub:def}

\begin{figure}[ht]
\centering
%%\begin{tikzpicture}[scale=1.0, every node/.style={transform shape}]
%\begin{tikzpicture}[scale=1.0]
%\tikzset{
%    very thin/.style=  {line width=0.5pt},
%    thin/.style=       {line width=.6pt},
%    ultra thick/.style={line width=2pt}
%}
%\draw [thin] (0,0) to[short,*-] (1.3,0);
%\draw [thin] (6.0,0) to[short,*-] (4.7,0);
%\draw [thin] (1.3,0.9) -- (4.7,0.9) -- (4.7,-0.9) -- (1.3,-0.9) -- (1.3,0.9);
%\draw [-{latex}, ultra thick] (-0.8,0) -- (-0.1,0);
%\draw [-{latex}, ultra thick] (6.8,0) -- (6.1,0);
%\draw [-{latex}, very thin] (0,-0.2) -- (0,-1) -- (0.5,-1);
%\draw [-{latex}, very thin] (6,-0.2) -- (6,-1) -- (6.5,-1);
%\node at (0.25,-1.4) {$x_2$};
%\node at (6.25,-1.4) {$x_1$};
%\node at (6.6,0.4) {$F$};
%\node at (-0.6,0.4) {$F$};
%\node at (3,0.3) {\large Mechanical};
%\node at (3,-0.2) {\large one-port};
%\end{tikzpicture}
%%\end{center}
\includegraphics[totalheight=3cm]{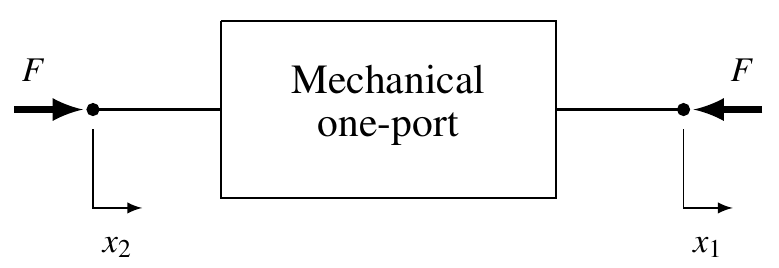}
    \caption{A free-body diagram of a one-port (two-terminal)
      mechanical element or network with force-velocity pair $(F,v)$
      where $v = \dot{x}$, $x = {x}_2 - {x}_1$.}
    \label{fig:mechport}
\end{figure}

We will consider (idealised) mechanical elements or networks which take the form of a {\em mechanical one-port} as shown  
in Fig.~\ref{fig:mechport}.  The one-port has two {\em terminals} for connection to other elements or networks.  The terminals are subject to an equal and opposite force $F$ 
and have absolute displacements $x_1$ and $x_2$.  Fig.~\ref{fig:mechport} illustrates the sign convention whereby a positive $F$ corresponds to a compressive force and a positive $x = x_2 - x_1$ corresponds to the terminals moving towards each other.  The force $F$ is an example of a {\em through-variable} and the relative displacement $x$ (and relative velocity $\dot{x}$ and relative acceleration $\ddot{x}$) is an {\em across-variable} \cite{shearer1967introduction}.  Either or neither of the variables may be considered an ``input''.  A {\em device law} for a mechanical one-port is a relation between through- and across-variables. 

The three elementary linear, passive, time-invariant mechanical one-port elements with two independently moveable terminals are the spring, damper and inerter  \cite{smith2002synthesis} with ideal modelling equations:  $F(t)=kx(t)$, $F(t)=c\dot{x}(t)$ and $F(t)=b\ddot{x}(t)$ where $k>0$, $c>0$ and $b>0$.  In the force-current  mechanical-electrical analogy these elements are analogous to the inductor, resistor and capacitor respectively.  The mass element may also be considered to be a one-port as in Fig.~\ref{fig:mechport} with the mass being rigidly attached to terminal two, and terminal one being a fixed point in the inertial frame of reference (see \cite{shearer1967introduction}), and as such is analogous to a grounded capacitor.  It is implicit in the definition of the spring, damper and inerter that they have negligible mass compared to any masses to which they are connected.

Mechanical one-ports in rotary form have two terminals which are independently rotatable about a common axis.  The rotary spring, damper and inerter \cite{smith2001force} are characterised by the equal and opposite torque applied to the terminals being proportional to the relative angular displacement, velocity and acceleration between the terminals respectively.  A pure inertia on a rotating shaft has only one terminal that can be rotated independently and, like the mass, is analogous to a grounded capacitor.  

\subsection{Adjustable one-ports}

We define a one-port to be {\em adjustable} if it has a parameter which may be freely varied as a function of time.  Such a parameter is considered to be a manipulable {\em input} to the device.  It allows the device to become part of a control system in which the parameter is adjusted by scheduling or feedback.   A ubiquitous example of an adjustable one-port is the variable damper defined by:
\begin{equation}
F(t) = c(t) \dot{x}(t), \hspace*{3mm} 0 < c_{\min} \leq c(t) \leq c_{\max}
\label{eq:semidamper}
\end{equation}
where $c(t)$ is the adjustable damper rate, and $c_{\min}$, $c_{\max}$ are minimum and maximum allowed values.  

\subsection{Passivity and losslessness}

The device laws we shall consider in this paper may be written in differential form:
\begin{equation}
h \left( F(t),\dot{F}(t),\ldots , x(t), \dot{x}(t), \ldots, u(t), \dot{u}(t), \ldots \right) = 0
\label{eq:devicelaw}
\end{equation}
for some function $h$, where $F(t)$ and $x(t)$ are defined in Section~\ref{sub:def} and $u(t)$ is a manipulable input.  We consider the set of (locally integrable, weak) solutions to (\ref{eq:devicelaw}) as the {\em behaviour} of the device in the sense of Willems \cite{willems2007behavioral}, \cite{polderman1998introduction}:
\begin{equation}
\mathcal{B} = \left\{ \left( F(t), x(t), u(t) \right) \in \mathcal{L}^{\mbox{\footnotesize loc}}_1 (\mathbb{R},\mathbb{R}^3) \mbox{ s.t. } (\ref{eq:devicelaw}) \mbox{ holds} \right\} ,
\end{equation}
where $\mathcal{L}^{\mbox{\footnotesize loc}}_1 (\mathbb{R},\mathbb{R}^3)$ denotes the functions from $\mathbb{R}$ into $\mathbb{R}^3$ that are Lebesgue integrable on any finite interval.  We adopt a definition of passivity from \cite{willems2007dissipative}, \cite{hughes2017theory}.

\begin{definition}\label{def:passive}
%\noindent {\bf Definition 1}.  
The device law (\ref{eq:devicelaw}) is {\em passive} if, for any $\left( F(t), x(t), u(t) \right) \in \mathcal{B}$ and $t_0 \in \mathbb{R}$, there exists $K \in  \mathbb{R}$ such that, 
\[
 \int_{t_0}^{t_1} {F}(t) \dot{x}(t) \, dt \geq -K
\]
for all $t_1 \geq t_0$, and where $K$ is independent of $\left\{ \left( {F}(t), {x}(t), {u}(t) \right) | \; t > t_0 \right\}$.
\end{definition}

%\begin{definition}\label{def:passive}
%%\noindent {\bf Definition 1}.  
%The device law (\ref{eq:devicelaw}) is {\em passive} if, for any $\left( F(t), x(t), u(t) \right) \in \mathcal{B}$ and $t_0 \in \mathbb{R}$, there exists $K \in  \mathbb{R}$ such that, 
%\[
%\int_{t_0}^{t_1} {F}(t) \dot{x}(t) \, dt \geq -K
%\]
%for all $t_1 \geq t_0$, where $K$ may depend on $t_0$ for any $\left( {F}_1(t), {x}_1(t), {u}_1(t) \right) \in \mathcal{B}$ for which $\left( F(t), x(t), u(t) \right) = \left( {F}_1(t), {x}_1(t), {u}_1(t) \right)$ for all $t \leq t_0$.
%\end{definition}
%

As noted in \cite{willems2007dissipative} the definition expresses the fact that,
for any trajectory, and starting at any particular
time, the net amount of energy that may be extracted subsequently from the
device cannot be arbitrarily large, namely
\[
- \int_{t_0}^{t_1} {F}(t) \dot{x}(t) \, dt \leq K .
\]
The provision that $K$ may depend on $t_0$ and on the trajectory prior to $t_0$, but must be independent of possible future trajectories, is included following \cite{hughes2017theory}, where its importance is illustrated by \cite[Example 6]{hughes2017theory} in time-varying and non-linear cases.

In \cite{hughes2016controllability} single-input single-output systems are considered whose behaviour is defined by the solutions to the equation
\[
p(\frac{d}{dt}) F(t) = q(\frac{d}{dt}) v(t)
\]
where $p$ and $q$ are real polynomials (where $v(t) = \dot{x}(t)$).  It is shown that the behaviour is passive if and only if $g(s) = p(s)/q(s)$ is a positive-real function and $p(s)$ and $q(s)$ have no common roots in the closed right half plane, unless $g(s) + g(-s) = 0$ in which case $p(s)$ and $q(s)$ are coprime.  (See \cite{hughes2017theory} for the generalisation to the multi-port case.)

It may be observed that the device law (\ref{eq:semidamper}) satisfies
\[
\int_{t_0}^{t_1} {F}(t) \dot{x}(t) \, dt = \int_{t_0}^{t_1} {c}(t) \dot{x}(t)^2 \, dt \geq 0 
\]
for any $t_0 \leq t_1$, and hence such devices are passive in a formal sense.  Sometimes the terminology ``semi-active'' is used in the literature since a small amount of power may be required in practice to make the adjustments.  Our approach in this paper is to classify devices as passive in terms of their terminal behaviour according to Definition~\ref{def:passive}, or if not, to refer to them as {\em active}.   
We now follow \cite{hughes2016controllability} in defining losslessness.

\begin{definition}\label{def:lossless}
The device law (\ref{eq:devicelaw}) is {\em lossless} if it is passive and if, for any $\left( F(t), x(t), u(t) \right) \in \mathcal{B}$ and $t_0, t_1 \in \mathbb{R}$
\[
\int_{t_0}^{t_1} {F}(t) \dot{x}(t) \, dt = 0
\]
whenever $F(t)$, $x(t)$, $u(t)$ and all derivatives are equal at $t_0$ and $t_1$.
\end{definition}

The above definition states simply that, in addition to being passive, there must be zero net energy transfer to or from the device over a time interval whenever the initial and final conditions are identical.  Springs and inerters are lossless according to this definition.  Our goal in this paper is to determine if springs and inerters may be adjustable as well as lossless.  In the first instance this question may be addressed purely in terms of candidate device laws.  There is then a further question as to physical realisability.  Ordinary springs and inerters are realisable physically without a power supply, and it is clearly important to know if the same is true for any passive or lossless, adjustable device laws.

\section{Device laws}\label{sec:devicelaws}

In this section we discuss some candidate device laws for adjustable springs and inerters in general terms, without considering the question of physical realisability.

\subsection{Device laws for adjustable springs}

\begin{example}\label{ex:1}
%\noindent {\bf Example 1} 
(directly adjustable spring constant). Let
\begin{equation}
F(t) = k(t) x(t).\label{eq:directadjust}
\end{equation}
This is the mostly commonly assumed device law for a ``semi-active'' (i.e. passive) spring.  It is in fact active (this fact is also pointed out in \cite{garrido2018assumed}).  Assuming that $k(t_1) = k(t_0)$ and $x(t_1) = x(t_0)$ we have:
\begin{eqnarray*}
\mathcal{E} & = & \int_{t_0}^{t_1} F(t) \dot{x}(t) \, dt \\
& = &  \int_{t_0}^{t_1} k(t) \frac{d}{dt} \left( \frac{1}{2} x^2(t) \right)\, dt  \\
& = &  - \frac{1}{2}  \int_{t_0}^{t_1} \dot{k}(t)  x^2(t) \, dt  .
\end{eqnarray*}
{\blue A trajectory can be constructed} for which $\mathcal{E}$ is negative.  For example, with $t_0=0$, $t_1 = 4$, $x(t) = t$ for $0\leq t \leq 2$, $x(t) = 4-t$ for $2\leq t \leq 4$, $k(0)=2$, $\dot{k}(t) = -1$ for $0\leq t \leq 1$, $\dot{k}(t) = 1$ for $2\leq t \leq 3$, and $\dot{k}(t) = 0 $ otherwise, we find that $\mathcal{E} = -1$.  Hence, if such a cycle is continually repeated, an arbitrary amount of energy can be extracted, namely there is no $K \in  \mathbb{R}$ for which the conditions of Definition~\ref{def:passive} hold. 
\end{example}

\begin{example}\label{ex:2}
%\noindent {\bf Example 2}
 (adjustable spring constant with smoothing). Let
\begin{equation}
F(t) = k(t) x(t) -  \int_{-\infty}^t \dot{k}(\tau)  x(\tau) \, d\tau. \label{eq:withsmoothing}
\end{equation}
The above is an idealised device law inspired by the behaviour of the mechanism of Fig.~1 in \cite{jabbari2002vibration} for a step increase in $k(t)$ (but not a decrease).  Differentiating (\ref{eq:withsmoothing}) gives $\dot{F}(t) = k(t) \dot{x}(t)$.  Hence, assuming that $k(t_1) = k(t_0)$, $x(t) = x(t_0)$ and $F(t_1) = F(t_0)$ we have:
\begin{eqnarray*}
\mathcal{E} & = & \int_{t_0}^{t_1} F(t) \dot{x}(t) \, dt \\
& = & - \int_{t_0}^{t_1} \dot{F}(t) {x}(t) \, dt \\
& = &  - \int_{t_0}^{t_1} k(t) \frac{d}{dt} \left( \frac{1}{2} x^2(t) \right)\, dt  \\
%& = &  \int_{t_0}^{t_1} k(t) \frac{d}{dt} \left( \frac{1}{2} x^2(t) \right)\, dt  \\
& = &   \frac{1}{2}  \int_{t_0}^{t_1} \dot{k}(t)  x^2(t) \, dt  .
\end{eqnarray*}
{\blue A trajectory can be constructed} for which $\mathcal{E}$ is negative.  For example, with $t_0=0$, $t_1 = 6$, $x(t) = t$ for $0\leq t \leq 3$, $x(t) = 6-t$ for $3\leq t \leq 6$, $k(0)=2$, $\dot{k}(t) = -1/2$ for $0\leq t \leq 1$ and $2\leq t \leq3$, $\dot{k}(t) = 1$ for $4\leq t \leq 5$, and $\dot{k}(t) = 0 $ otherwise, we find that 
\[
\int_{t_0}^{t_1} k(t) \dot{x}(t) = 0
\]
which implies $F(t_1) = F(t_0)$.  Furthermore $\mathcal{E} = -1$.  Hence, if such a cycle is continually repeated, an arbitrary amount of energy can be extracted.  Hence the conditions of Definition~\ref{def:passive} cannot be satisfied, 
and the device law is active.
\end{example}

\begin{example}
%\noindent {\bf Example 3} 
(adjustable spring constant with up-smoothing). Let
\begin{equation}
F(t) = k(t) x(t) -  \int_{-\infty}^t {( \dot{k}(\tau) )}_+  x(\tau) \, d\tau. \label{eq:withupsmoothing}
\end{equation}
where
\[
 \left( u(t) \right)_+ = \left\{ \begin{array}{cl}
 u(t) & \mbox{ when $u(t) \geq 0$} \\
 0 & \mbox{ otherwise}
 \end{array}\right.
\]
with $ \left( u(t) \right)_-$ defined similarly, so that $u(t) = \left( u(t) \right)_+ + \left( u(t) \right)_-$.
The above idealised device law is a continuous version of the mechanism of Fig.~1 in \cite{jabbari2002vibration} for increasing $k(t)$ and corresponds to Example 1 otherwise.  Differentiating (\ref{eq:withupsmoothing}) gives 
\[
\dot{F}(t) = k(t) \dot{x}(t) + {( \dot{k}(t) )} _- x(t).
\]
Hence, assuming that $k(t_1) = k(t_0)$, $x(t_1) = x(t_0)$ and $F(t_1) = F(t_0)$ we have:
\begin{eqnarray*}
\mathcal{E} & = & \int_{t_0}^{t_1} F(t) \dot{x}(t) \, dt \\
& = & - \int_{t_0}^{t_1} \dot{F}(t) {x}(t) \, dt \\
%& = &  - \int_{t_0}^{t_1} k(t) \frac{d}{dt} \left( \frac{1}{2} x^2(t) \right)\, dt  \\
%& = &  \int_{t_0}^{t_1} k(t) \frac{d}{dt} \left( \frac{1}{2} x^2(t) \right)\, dt  \\
& = &   \frac{1}{2}  \int_{t_0}^{t_1} \dot{k}(t)  x^2(t) \, dt   -   \int_{t_0}^{t_1} {( \dot{k}(t) )}_-  x^2(t) \, dt \\
& = &   \frac{1}{2}  \int_{t_0}^{t_1} \left( {( \dot{k}(t) )}_+ - {( \dot{k}(t) )}_- \right) x^2(t) \, dt \\
& = &   \frac{1}{2}  \int_{t_0}^{t_1} |\dot{k}(t)|  x^2(t) \, dt  > 0.
\end{eqnarray*}
Here $\mathcal{E}$ is always positive, so energy cannot be extracted over a repeating cycle.  This doesn't yet show that the device law is passive, though clearly it cannot be lossless.  In fact, it fails also to be passive.  Let $t_0 = 0$, $t_1 = 2n$ for some positive integer $n$ and suppose $k=2$ for $t < 0$,
\begin{eqnarray*}
& & k=2 + \sin (2\pi t ), \; x(t) = 1, \; (0 \leq t \leq n), \\
&& k = 2, \; x(t) = t+1-n, \; (n \leq t \leq 2n).
\end{eqnarray*}
From (\ref{eq:withupsmoothing}) we find that $F(n) = 2-2n$, which means $F(t) = 2x(t) - 2n$ for $n \leq t \leq 2n$. Hence 
\[
\int_{0}^{2n} {F}(t) \dot{x}(t) \, dt = 2n - n^2 
\]
which cannot be bounded below independent of $n$.  Hence Definition~\ref{def:passive} cannot be satisfied.
\end{example}

\begin{example}\label{ex:4}
%\noindent {\bf Example 4} 
(adjustable spring constant with semi-smoothing). Let
\begin{equation}
F(t) = k(t) x(t) -  \frac{1}{2} \int_{-\infty}^t \dot{k}(\tau)  x(\tau) \, d\tau. \label{eq:withsymsmoothing}
\end{equation}
From Examples 1 and 2, over any cycle in which $k(t_1) = k(t_0)$, $x(t_1) = x(t_0)$ and $F(t_1) = F(t_0)$ we have 
\[
%\mathcal{E} = 
\int_{t_0}^{t_1} F(t) \dot{x}(t) \, dt  = 0.
\]
Evidently this law has the potential to be lossless, however we now show that it fails to be so since it is not passive.  We first note that
\[
\dot{F}(t) = k(t) \dot{x}(t) + \frac{1}{2} \dot{k}(t) x(t).
\]
It now follows that
\begin{eqnarray*}
%\mathcal{E} 
 \int_{t_0}^{t_1} F(t) \dot{x}(t) \, dt 
& = & F(t_1) x(t_1)  -F(t_0) x(t_0) -  \int_{t_0}^{t_1} \dot{F}(t) {x}(t) \, dt \\
& = & F(t_1) x(t_1)  -F(t_0) x(t_0)  - \frac{1}{2}  \int_{t_0}^{t_1} \dot{k}(t)  x^2(t) \, dt  \\
&  &   \hspace*{2mm}  - \int_{t_0}^{t_1} k(t) \frac{d}{dt} \left( \frac{1}{2} x^2(t) \right)\, dt  \\
& = & F(t_1) x(t_1) - \frac{1}{2} k(t_1) x(t_1)^2 \\
& & \hspace*{2mm} -F(t_0) x(t_0) 
+ \frac{1}{2} k(t_0) x(t_0)^2  .
\end{eqnarray*}
Let $t_0 = 0$, $t_1 = n + \frac{3}{4}$ for some positive integer $n$, suppose $k=3$ for $t < 0$ and 
\begin{eqnarray*}
& & k=2 + \cos (2\pi t ), \; x(t) = \sin (2\pi t ), \; (t > 0). 
\end{eqnarray*}
Then 
$x(t_1) =-1$, $k(t_1) =2$, and from (\ref{eq:withsymsmoothing}) 
$F(t_1) = \frac{(4n+3)\pi}{8} - 2$.  Hence
\[
\int_{t_0}^{t_1} {F}(t) \dot{x}(t) \, dt =  -\frac{(4n+3)\pi}{8} + 1
\]
which cannot be bounded below independent of $n$.  Hence Definition~\ref{def:passive} cannot be satisfied.
\end{example}

\subsection{Device laws for adjustable inerters}

\begin{example}
%\noindent {\bf Example 5} 
(directly adjustable inertance). Let
\begin{equation}
F(t) = b(t) \ddot{x}(t).\label{eq:directadjustinertance}
\end{equation}
This is the mostly commonly assumed device law for a ``semi-active'' inerter.  It is again active.  Assuming that $b(t_1) = b(t_0)$ and $\dot{x}(t_1) = \dot{x}(t_0)$ we have:
\begin{eqnarray*}
\mathcal{E} & = & \int_{t_0}^{t_1} F(t) \dot{x}(t) \, dt \\
& = &  \int_{t_0}^{t_1} b(t) \frac{d}{dt} \left( \frac{1}{2} \dot{x}^2(t) \right)\, dt  \\
& = &  - \frac{1}{2}  \int_{t_0}^{t_1} \dot{b}(t)  \dot{x}^2(t) \, dt . 
\end{eqnarray*}
{\blue A trajectory can be constructed} for which $\mathcal{E}$ is negative, 
e.g.\ with $b(t)$ and $\dot{x}(t)$ chosen as $k(t)$ and $x(t)$ in Example~\ref{ex:1}.  
Such a device could be operated in a repeating cycle which extracts a net amount of energy in each cycle.
 Hence Definition~\ref{def:passive} cannot be satisfied.
\end{example}

\begin{example}
%\noindent {\bf Example 6} 
(inerter with actively controlled fly-weights). Let
\begin{equation}
F(t) =  \frac{d}{dt} \left( b(t) \dot{x}(t) \right) .\label{eq:directadjustinertance6}
\end{equation}
Such a device is described in \cite{garrido2018assumed}---a rack and pinion is used to convert linear motion into the rotary motion of two arms with weights which are moved in or out by actuators.  It is shown that (\ref{eq:directadjustinertance6}) holds for the device.
Hence, assuming that $b(t_1) = b(t_0)$ and $\dot{x}(t) = \dot{x}(t_0)$ we have:
\begin{eqnarray*}
\mathcal{E} & = & \int_{t_0}^{t_1} F(t) \dot{x}(t) \, dt \\
& = &  - \int_{t_0}^{t_1} b(t) \frac{d}{dt} \left( \frac{1}{2} \dot{x}^2(t) \right)\, dt  \\
%& = &  - \int_{t_0}^{t_1} k(t) \frac{d}{dt} \left( \frac{1}{2} x^2(t) \right)\, dt  \\
%& = &  \int_{t_0}^{t_1} k(t) \frac{d}{dt} \left( \frac{1}{2} x^2(t) \right)\, dt  \\
& = &   \frac{1}{2}  \int_{t_0}^{t_1} \dot{b}(t)  \dot{x}^2(t) \, dt  .
\end{eqnarray*}
Again $\mathcal{E}$ can be negative, 
e.g.\ with $b(t)$ and $\dot{x}(t)$ chosen as $k(t)$ and $x(t)$ in Example~\ref{ex:2}.
Hence Definition~\ref{def:passive} cannot be satisfied,
so the device law is active, as pointed out in \cite{garrido2018assumed}. 
\end{example}

\section{Planar mechanism for lossless adjustable devices}\label{sec:planar}

\subsection{Lossless adjustable spring}\label{sec:lossadjspr}

\begin{figure}[ht] % SPRING and LEVER
\centering
\includegraphics[totalheight=5cm]{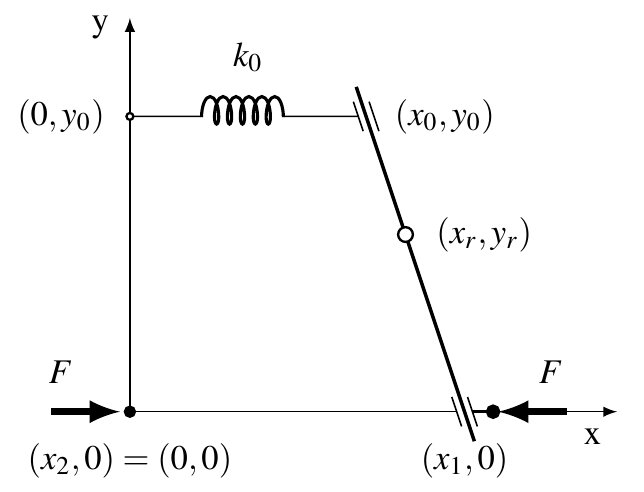}
\caption{Spring and lever with moveable fulcrum at $(x_{r},y_r)$.}
\label{fig:planarmech}
\end{figure}

We consider a theoretical mechanism as depicted in Fig.~\ref{fig:planarmech} in which the x- and y-axes are 
fixed in the device housing.  
The device terminals are located at $(0,0)$ and $(x_1,0)$ according to the convention of Fig.~\ref{fig:mechport} and accordingly we define $x=-x_1$ and $v=-\dot{x}_1$.  An internal spring with stiffness $k_0$ is constrained to move parallel to the x-axis with fixed y-coordinate $y_0$ and generates a force equal to $-k_0x_0$.  An ideal massless lever has a moveable fulcrum at $(x_r,y_r)$. 

Taking moments about the fulcrum gives: 
\[
Fy_r = k_0 x_0 (y_0 - y_r).
\]
The geometrical position of the lever imposes the following constraint:
\[
\frac{x_0-x_1}{y_0} = \frac{x_r - x_1}{y_r}
\]
which, after replacing $x_1$ by $-x$, can be written as:
\begin{equation}
x_0 = rx +(r+1)x_r \label{eq:x0}
\end{equation}
where 
\begin{equation}
r = (y_0 - y_r)/y_r.\label{eq:rdef}
\end{equation}  
We therefore obtain
\begin{equation}
F = k_0 r^2 x + k_0 r (r+1) x_r.  \label{eq:F11}
\end{equation}

We now consider the fulcrum to be moveable with an imposed condition that the instantaneous power supplied at the external terminals of the device equals the rate of change of the internal energy of the spring.  This is equivalent to $F\dot{x} = k_0 x_0 \dot{x}_0$, which on noting that $F=k_0x_0 r$ is in turn equivalent to $r\dot{x} = \dot{x}_0$.  The latter, using \eqref{eq:x0}, is equivalent to:
\begin{equation}
(r+1)\dot{x}_r + \dot{r} (x + x_r) = 0. \label{eq:rdot}
\end{equation}
{\blue Absorbing $k_0$ into $r$ (or equivalently setting $k_0=1$)} and making the substitution $w=(r+1) x_r$ we can reduce (\ref{eq:F11}) and (\ref{eq:rdot}) to the form:
\begin{eqnarray}
F & = & r^2 x + r w, \label{eq:la1}\\
\dot{w} & = & -\dot{r} x. \label{eq:la2}
\end{eqnarray}

Equations (\ref{eq:la1}), (\ref{eq:la2}) can be written in an integral form as a single equation with the substitution $k(t) = r(t)^2$:
\begin{equation}
F(t) = k(t) x(t) -  \frac{1}{2} k(t)^{1/2} \int_{-\infty}^{t}  k(\tau)^{-1/2} \dot{k}(\tau)  x(\tau) \, d\tau. 
%F = k x -  \frac{1}{2} k^{1/2} \int_{t_0}^{t_1}  k()^{-1/2} \dot{k}(\tau)  x(\tau) \, d\tau. 
\label{eq:squarerootform}
\end{equation}
It is interesting to note the close similarity with the device law (\ref{eq:withsymsmoothing}) of Example~\ref{ex:4}.

Further from (\ref{eq:la1}), (\ref{eq:la2}) we define
\[
\mathcal{I} = \frac{1}{2} (xr +w)^2 = \frac12 \frac{F^2}{r^2} .
\]
and note that
\begin{eqnarray*}
\frac{d}{dt} \left( \mathcal{I} \right) & = & (xr+w) (\dot{x}r + x\dot{r} + \dot{w}) \\
& = & (xr+w) \dot{x} r = F\dot{x} .
\end{eqnarray*}
Hence $\mathcal{I}$ can be considered to be the internal stored energy of the device law (\ref{eq:la1})--(\ref{eq:la2}) and it follows that the device law is lossless according to Definition~\ref{def:lossless}.  Note that we may take $K= \mathcal{I}(t_0)$ to show passivity according to Definition~\ref{def:passive}.

\subsection{Lossless adjustable inerter}\label{sub:adjinerter}

\begin{figure}[ht] % INERTER
\centering
\includegraphics[totalheight=5cm]{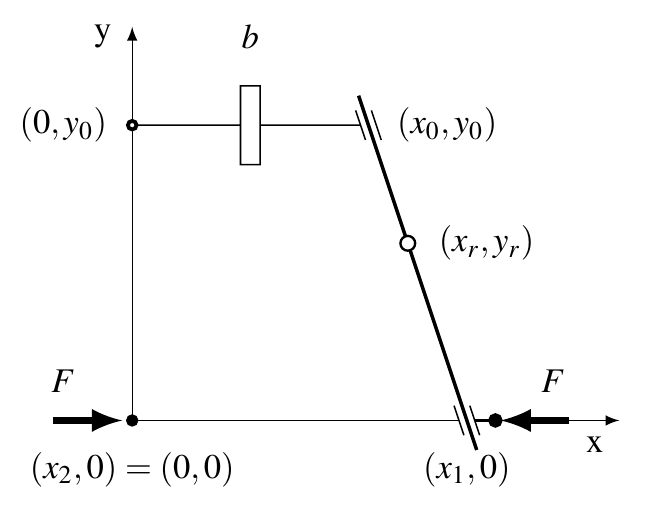}
\caption{Inerter and lever with moveable fulcrum at $(x_{r},y_r)$.}
\label{fig:planarinerter}
\end{figure}

We consider the mechanism as depicted in Fig.~\ref{fig:planarinerter} which is similar to the device in  Fig.~\ref{fig:planarmech} except that the spring is replaced by an inerter which generates a force equal to $-b \ddot{x}_0$. 

Taking moments about the fulcrum gives $F=b\ddot{x}_0r$ with $r$ defined as in (\ref{eq:rdef}) and $x_0$ as in (\ref{eq:x0}).  Applying again the condition that the instantaneous power supplied at the external terminals of the device equals the rate of change of the internal energy of the inerter gives again (\ref{eq:rdot}) or equivalently $\dot{x}_0 = r\dot{x}$.  {\blue Absorbing $b$ into $r$ (or equivalently setting $b=1$)} we obtain the following form for the device law:
\begin{equation}
F = r \frac{d}{dt} \left( r\dot{x} \right) . \label{eq:inerterDL}
\end{equation}
It is immediate to see that $F\dot{x} = \frac{d}{dt} \left( \mathcal{I} \right)$ where we may define the internal stored energy by:
\[
\mathcal{I} = \frac{1}{2} r^2\dot{x}^2.
\]
Hence the device law (\ref{eq:inerterDL}) is lossless according to Definition~\ref{def:lossless}.  

\subsection{Dual canonical form for the lossless adjustable spring}\label{sub:dualspring}

The simplicity of the device law (\ref{eq:inerterDL}) in the inerter case is in striking contrast to (\ref{eq:la1}--\ref{eq:la2}) for the spring.  We will now show that (\ref{eq:la1}--\ref{eq:la2}) can be rewritten in a dual form to (\ref{eq:inerterDL}).  Differentiating (\ref{eq:la1}) and making use of (\ref{eq:la2}) we have:
\begin{eqnarray*}
\dot{F} & = & 2r\dot{r} x + r^2 \dot{x} + r\dot{w} + \dot{r} w \\
& = &  r^2 \dot{x} + \frac{\dot{r}}{r} F .
\end{eqnarray*}
Writing $p=r^{-1}$ we deduce that
\begin{equation}
\dot{x} = p \frac{d}{dt} \left( pF \right) . \label{eq:springDL}
\end{equation}
Again it is immediate to see that $F\dot{x} = \frac{d}{dt} \left( \mathcal{I} \right)$ where we may define the internal stored energy by:
\[
{
\mathcal{I} = \frac{1}{2} p^2 F^2.
}
\]
Hence the device law (\ref{eq:springDL}) is lossless according to Definition~\ref{def:lossless}.  

\section{Canonical device laws}\label{sec:canonical} 

The device laws (\ref{eq:inerterDL}) and (\ref{eq:springDL}) have been shown to be lossless according to Definition~\ref{def:lossless}.  This does not as yet show that they may be realised physically without the need for an internal power source.  We consider this next.

\subsection{Physical implementation}\label{sub:phys}

For Fig.~\ref{fig:planarmech} or Fig.~\ref{fig:planarinerter} the condition that the instantaneous power supplied at the external terminals of the device equals the rate of change of the internal energy of the spring or inerter reduces to the same equation (\ref{eq:rdot}).  This determines the manner in which the fulcrum should be moved when the ratio $r$ is changed.  {\blue The condition ensures that the reaction forces at the fulcrum are constrained to do no work.}
We now examine this condition further.  Eliminating $r$ using (\ref{eq:rdef}) then (\ref{eq:rdot}) reduces to
\[
0 = y_r \dot{x}_r - \dot{y}_r (x_r - x_1)
\]
which means geometrically that the vectors 
\[
\left(
\begin{array}{c}
\dot{x}_r \\
\dot{y}_r 
\end{array}
\right)
\mbox{ and }
\left(
\begin{array}{c}
{x}_r - x_1 \\
{y}_r 
\end{array}
\right)
\]
are parallel.  The fulcrum must always move parallel to the bar. 

\begin{figure}[ht] % TABLE TOP CONFIGURATION
\centering
\includegraphics[totalheight=3cm]{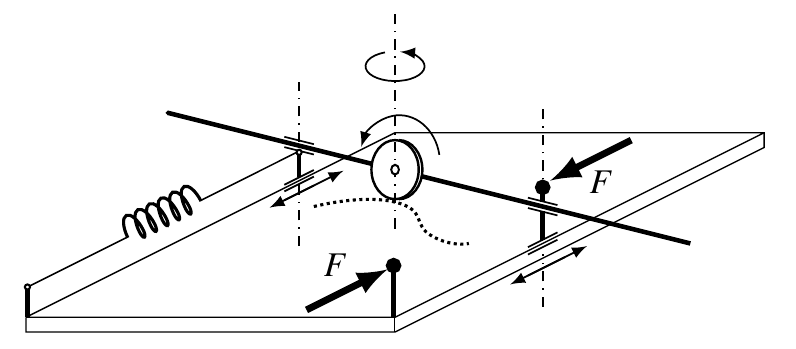}
\caption{Schematic of a lever mechanism with moveable fulcrum to allow a physical realisation of lossless adjustable springs and inerters.}
\label{fig:fulcrum}
\end{figure}

A conceptual scheme to realise such adjustability is shown in Fig.~\ref{fig:fulcrum}.  A wheel is attached to the bar at the fulcrum and is free to rotate about a vertical axis through the fulcrum and the contact point of the wheel on a supporting table.  The wheel is allowed to rotate about a horizontal axis which is perpendicular to the bar to produce a rolling motion on the table which is always instantaneously parallel to the bar.  The rolling of the wheel is the means of mechanism adjustment by altering the ratio $r$ or $p=r^{-1}$ with $r$ defined as in (\ref{eq:rdef}). 

%%%%%%%%%%%%%%%%%%%%%%%%%%%%%%%%%

{\red We remark that recent examples of moveable pivot \cite{jafari2011awas,groothuis2014variable,liu2018modeling,lazarek2018design} for the purpose of adjusting variable stiffness involve a predetermined motion path for the pivot, typically a straight line, which will not satisfy the above geometrical relations in all dynamic situations. Hence, such devices will not be able to implement a {\em lossless} adjustable spring or inerter.}

\subsection{The varspring and varinerter}

{\red Based on the construction of Fig.~\ref{fig:fulcrum},
it} appears justified to introduce a pair of ideal, lossless adjustable mechanical one-ports which we will name the {\em varspring} and {\em varinerter}.    The ideal devices are defined by the laws:
\begin{eqnarray}
 v & = & p \frac{d}{dt} \left( pF \right) \hspace*{7mm} \mbox{(varspring)} \label{eq:vars}\\
 F & = & r \frac{d}{dt} \left( rv \right) \hspace*{9mm} \mbox{(varinerter)}  \label{eq:vari}
\end{eqnarray}
%where $(F,v)$ is the force-velocity pair of the mechanical one-port and $p(t)$, $r(t)$ are positive and freely adjustable parameters.  The internal energy of the devices is given by $\frac{1}{2} p^2F^2$ and $\frac{1}{2} r^2v^2$ respectively.  As in the case of the ideal inerter \cite{smith2002synthesis} it is important that physical devices may be constructed which approximate the ideal behaviour, have two terminals which are freely and independently moveable,  have a sufficiently small mass and sufficient travel, and do not need to have a physical attachment to a fixed point in space (see Section II.C in \cite{smith2002synthesis}). The construction of Fig.~\ref{fig:fulcrum} suggests that devices satisfying these conditions are physically realisable in principle.  The varinerter is realised as in Fig.~\ref{fig:fulcrum} with an inerter replacing the spring.
where $(F,v)$ is the force-velocity pair of the mechanical one-port and $p(t)$, $r(t)$ are positive and freely adjustable parameters.  The internal energy of the devices is given by $\frac{1}{2} p^2F^2$ and $\frac{1}{2} r^2v^2$ respectively.  
{\blue It is important that physical devices may be constructed which approximate the ideal behaviour, for example, having sufficiently small dissipation through friction, and as in the case of the ideal inerter \cite{smith2002synthesis}, sufficiently small mass, sufficient travel, have no physical attachment to a fixed point in space, and have two terminals which are freely and independently moveable (see Section II.C in \cite{smith2002synthesis}).} The construction of Fig.~\ref{fig:fulcrum} suggests that devices satisfying these conditions are physically realisable in principle.  The varinerter is realised as in Fig.~\ref{fig:fulcrum} with an inerter replacing the spring. {\blue We note that the above construction of the varspring and varinerter in Fig.~\ref{fig:fulcrum} can be conceptualized as a lossless adjustable two-port transformer with one of the ports terminated with either a spring or an inerter.}

It is interesting to note that the device laws (\ref{eq:vars}) and (\ref{eq:vari}) may be written in integral form as follows:
\begin{eqnarray}
 F(t) & = & \hat{r}(t) \int_{-\infty}^t  \hat{r}(\tau )v(\tau )  \, d\tau \hspace*{11mm} \mbox{(varspring)} \label{eq:varspring2}\\
 v(t) & = & \hat{p}(t) \int_{-\infty}^t  \hat{p}(\tau )F(\tau )  \, d\tau \hspace*{9mm} \mbox{(varinerter)}
\end{eqnarray}
where $\hat{r} = p^{-1}$ and $\hat{p} = r^{-1}$.  Note also that (\ref{eq:varspring2}) reduces to (\ref{eq:squarerootform}) with the substitution $k(t) = \hat{r}(t)^2$ and integration by parts.

\section{Rotary mechanical one-ports}\label{sec:rotary}

In this section we explore the rotary equivalents of the varspring and varinerter.  {\blue Motivated by the method of constructing the translational varspring and varinerter in Sections \ref{sec:lossadjspr} and \ref{sub:adjinerter} we first consider the possibility of an adjustable rotary transformer.}

\subsection{A lossless adjustable transformer}

\begin{figure}[ht] % ROTARY
\centering
\includegraphics[totalheight=5cm]{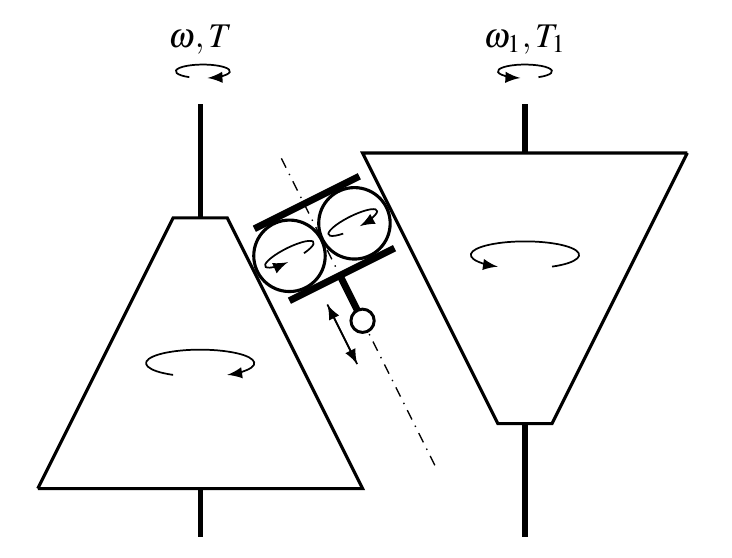}
\caption{Schematic of an adjustable rotary transformer with counter-rotating cones and continuously moveable connecting assembly \blue consisting of a pair of rotating balls within a housing.}
\label{fig:balls}
\end{figure}

We consider the construction depicted in Fig.~\ref{fig:balls} consisting of two right circular cones of equal aperture on parallel rotating shafts, with opposite orientation, and hence a constant perpendicular distance between the surfaces. Between the cones is an assembly consisting of two balls within a housing which is moveable parallel to the surface of the cones to maintain contact of the balls with the cones at the feet of the perpendicular beween the cones.  It is assumed that pure rolling is maintained between the balls and the cones, and between themselves, and that there is frictionless sliding between the balls and the housing.  With the assumption of negligible mass of the whole system the torques on the two shafts are proportional, with the proportionality being the instantaneous ratio of cone radii.  The assumption of pure rolling means that the angular velocities are similarly proportional.  Thus we may presume laws of the form:
\begin{eqnarray}
T_1 & = & p T, \label{eq:adjtran1} \\
\omega_1 & = & -p^{-1} \omega  \label{eq:adjtran2}
\end{eqnarray}
where $T$, $T_1$ are the torques on the shafts, $\omega$, $\omega_1$ are their angular velocities, and $p=p(t) > 0$ is the instantaneous ratio of cone radii.  We note that $T_1 \omega_1 + T \omega = 0$ so that no energy is absorbed or dissipated in the ideal device.  Hence we may consider the schematic of Fig.~\ref{fig:balls} as a physical realisation of a lossless adjustable rotary transformer.

{\red 
It is important to emphasize that, besides being lossless, an essential feature of the mechanism in Fig.~\ref{fig:balls} is that the ratio between angular velocities can be freely adjusted, including the case where the angular velocities are zero, as occurs when there is a reversal of sign. This feature contrasts with typical concepts of a continuously variable transmission (CVT), e.g., \cite{brokowski2002toward,rotella2018power}.

%It is important to emphasize that there are two essential features of the mechanism in Fig.~\ref{fig:balls}. First that it is lossless and second that it is capable of freely adjusting the ratio between angular velocities, including cases where the angular velocities experience reversal of sign. 

%EARLIER:\\
%The construction in Fig.~\ref{fig:balls} can be viewed as a model for a continuous variable transmission (CVT). It appears quite distinct from existing CVT concepts, in that change in the gearing ratio in Fig.~\ref{fig:balls} can be effected in the absence of motion of the shafts, in contrast to designs as in the rotational CVT in \cite[Fig.~2]{brokowski2002toward} where this is not possible. Whereas the idealized concept of a variable adjustable transformer in Fig.~\ref{fig:balls} is lossless, a rigorous study of power lossess  in existing CVT designs appears rather challenging \cite{rotella2018power}.
}

\subsection{The rotary varspring and varinerter}

We first consider attaching a rotary spring of rotational stiffness $k >0 $ (constant) to the second shaft in Fig.~\ref{fig:balls} defined by $T_1 = - k \theta_1$ where $\dot{\theta}_1 = \omega_1$.  A passive (lossless) rotary mechanical one-port is formed with the following relationship between the equal and opposite torque applied to the external (rotary) terminals $T$ and the relative angular velocity $\omega$ between the terminals:
$\omega = -p \omega_1 = p k^{-1}  \frac{d}{dt} \left( pT \right) $. 
%\begin{eqnarray*}
%\omega & = & -p \omega_1 \\
%& = & p k^{-1}  \frac{d}{dt} \left( pT \right) . 
%\end{eqnarray*}
Similarly, if a rotary inerter (see \cite{smith2001force}) with rotational inertance $b >0$, defined by $T_1 = - b \dot{\omega}_1$, is connected across the second shaft in Fig.~\ref{fig:balls}  a passive (lossless) rotary one-port is formed satisfying  
%\begin{eqnarray*}
$T = -b r \dot{\omega}_1 = b r  \frac{d}{dt} \left( r \omega \right) $.   The constants $k$ and $b$ can be absorbed into $p$ and $r$ respectively {\blue (or equivalently setting $k=1$ and $b=1$)}.  

This motivates the following definitions of the rotary varspring and varinerter:
 \begin{eqnarray}
 \omega & = & p \frac{d}{dt} \left( pT \right) \hspace*{8mm} \mbox{(rotary varspring)} \\
 T & = & r \frac{d}{dt} \left( r\omega \right) \hspace*{8.3mm} \mbox{(rotary varinerter)}
\end{eqnarray}
where $(T,\omega)$ is the torque-angular-velocity pair of the mechanical one-port and $p(t)$, $r(t)$ are positive and freely adjustable parameters.  The internal energy of the devices is given by $\frac{1}{2} p^2T^2$ and $\frac{1}{2} r^2\omega^2$ respectively. 

{\blue It is interesting to compare the embodiments presented for the translational and rotary varsprings and varinerters. A practical issue that arises with continuous operation of the translational devices, implemented in the manner of Fig.~\ref{fig:fulcrum}, is that the movement of the fulcrum in the $x$-direction may exceed the allowable travel. No such issue arises with the rotary devices.
}

\section{Adjustable Electrical devices}\label{sec:elec}

We turn now to the possibility of adjustable electrical devices.
The variable resistor with device law $v(t) = R(t) i(t)$, where $v(t)$ is the voltage across the device, $i(t)$ the current through, and $R(t) >0$ the variable resistance, is of course a ubiquitous device that is formally passive in the same way as the adjustable damper.  For the capacitor and inductor there are analogous issues to the mechanical case in constructing devices which are passive, lossless, as well as adjustable. 
We begin with some examples
that highlight how varying capacitance or inductance, directly, leads to active elements.

% with variable capacitance and inductance. Once again, depending on how the varying component is implemented, the device may be active or passive.\\

\begin{example}
%\noindent {\bf Example 7} 
(adjustable parallel-plate capacitor). 
Following \cite[Ch.\ 10]{feynman1979feynman} we consider a parallel-plate capacitor with a dielectric slab which can be inserted by varying amounts between the plates as shown in Fig.~\ref{fig:capacitor}.  The defining equation for the capacitance is $C(t):=q(t)/v(t)$, where $q$ is the charge on the plates and $v$ is the voltage between them, from which it follows that: 
\begin{equation}\label{eq:varcapacitance}
i(t) = \frac{d}{dt} \left( C(t) v(t)  \right) .
\end{equation}
It is shown in \cite[Ch.\ 10]{feynman1979feynman} that 
\[
C = \frac{\epsilon_0 b}{d} \left( \kappa x + a -x \right)
\]
where $\epsilon_0$ is the permittivity of empty space and $\kappa$ is the dielectric constant of the dielectric; it is assumed that the plates are rectangular of length $a$ and width $b$, are at a distance $d$ apart, and the dielectric is inserted by a distance $x= x(t)$. 

Now consider a time interval $[t_0,t_1]$ in which $C(t_0)=C(t_1)$ and $v(t_0)=v(t_1)$.  Then the energy supplied to the device
\begin{eqnarray*}
\mathcal{E}  = \int_{t_0}^{t_1} i(t)v(t)dt &=& \int_{t_0}^{t_1} v(t)\frac{d(C(t)v(t))}{dt} dt\\
&=&  - \int_{t_0}^{t_1} C(t)v(t)\dot v(t)dt \\
&=& \frac12 \int_{t_0}^{t_1} \dot C(t)v^2(t)dt.
\end{eqnarray*}
As before $\mathcal{E}$ can be negative, 
e.g.\ with $C(t)$ and $v(t)$ chosen as $k(t)$ and $x(t)$ in Example~\ref{ex:2}.
Hence Definition~\ref{def:passive} cannot be satisfied,
so the device law is active, as may also be expected since a force is required to move the dielectric slab.  A similar conclusion holds if the distance $d$ between the plates is varied.\\
\begin{figure}[ht]
%\centering
%%\begin{tikzpicture}[scale=1.0, every node/.style={transform shape}]
%\begin{tikzpicture}[scale=0.9]
%\tikzset{
%    semi thin/.style=  {line width=0.8pt},
%    thin/.style=       {line width=.6pt},
%    very thin/.style=  {line width=0.2pt},
%    thick/.style={line width=1pt}
%}
%\draw [thin] (0,2.2) -- (7,2.2) -- (7,2.6) -- (0,2.6) -- (0,2.2);
%\draw [thin] (2,1.2) -- (9,1.2) -- (9,2.1) -- (2,2.1) -- (2,1.2);
%\draw [thin] (0,.7) -- (7,.7) -- (7,1.1) -- (0,1.1) -- (0,.7);
%\draw [thin] (2,.5) -- (2,0);
%\draw [thin] (7,.5) -- (7,0);
%\draw [-{latex}, thin] (4,.25) -- (2,.25);
%\draw [-{latex}, thin] (5,.25) -- (7,.25);
%
%%\draw [-{latex}, very thin, dashed] (2,1.65) -- (1,1.65);
%% movement left-right-arrow
%
%%\draw[-{latex},  thin, dashed] (2.5-1.8, 1.65) -- (2.5-1.2, 1.65);
%%\draw[-{latex},  thin, dashed] (2.5-1.9, 1.65) -- (2.5-2.5, 1.65);
%
%\draw[-{latex},  thin, dashed] (6+2.5-1.8+2.3, 1.65+.7) -- (6+2.5-1.2+2.3, 1.65+.7);
%\draw[-{latex},  thin, dashed] (6+2.5-1.9+2.3, 1.65+.7) -- (6+2.5-2.5+2.3, 1.65+.7);
%
%
%\draw [-{latex}, semi thin] (7.9,0.7) -- (7.9,1.16);
%\draw [-{latex}, semi thin] (7.9,2.6) -- (7.9,2.14);
%\node at (8.1,0.6) {$d$};
%\node at (4.5,.25) {$x$};
%\node at (3.5,2.4) {Plate 1};
%\node at (3.5,0.9) {Plate 2};
%\node at (5.5,1.65) {Dielectric slab};
%\end{tikzpicture}
%%\end{center}
\centering
\includegraphics[totalheight=2.5cm]{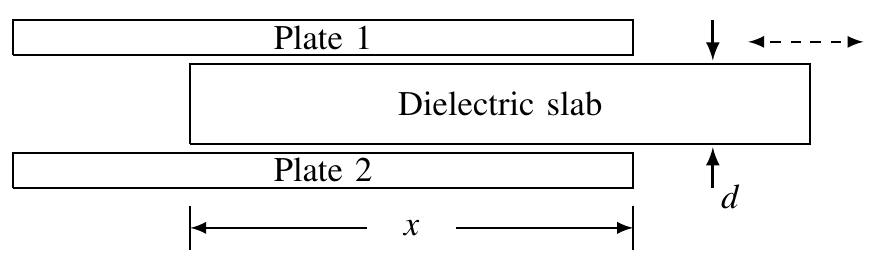}
    \caption{A parallel-plate capacitor with a moveable dielectric slab.}
    \label{fig:capacitor}
    \end{figure}
\end{example}

\begin{example}
%\noindent {\bf Example 8} 
(adjustable inductor with moveable core).  Likewise, the device law
\begin{equation}\label{eq:varinductance}
v(t)= \frac{d}{dt} \left( L(t) i(t) \right)  
\end{equation}
is the ideal law of a device whose inductance $L(t):=\Phi(t)/i(t)$ varies with time; here $\Phi$ is the magnetic flux through the coils, $i$ the current through, and $v$ the voltage across the terminals.  Such a variable inductor can be constructed with a moveable ferrite magnetic core as depicted in Fig.\ \ref{eq:variableinductance}, which is slid in or out of the coil in order to adjust the permeability and hence the magnetic flux.

\begin{figure}[ht]
\centering
\includegraphics[totalheight=4.5cm]{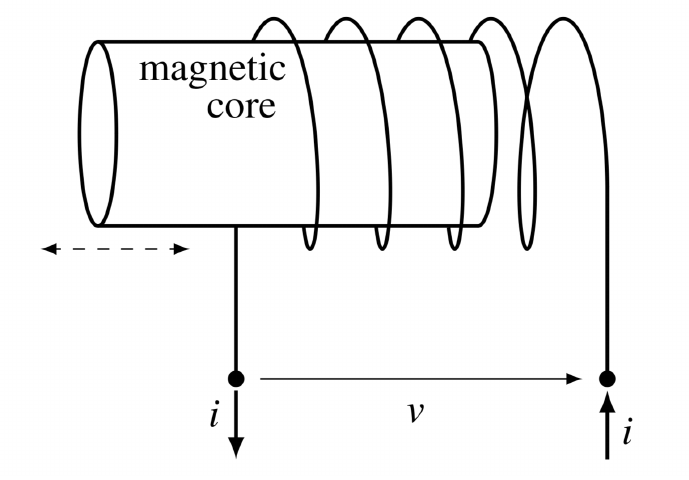}
\caption{Variable inductor with moveable magnetic core.}
\label{eq:variableinductance}
\end{figure}

Again consider a time interval $[t_0,t_1]$ in which $L(t_0)=L(t_1)$ and $i(t_0)=i(t_1)$.  Then the energy supplied to the device
\begin{eqnarray*}
\mathcal{E}  = \int_{t_0}^{t_1} i(t)v(t)dt &=& \int_{t_0}^{t_1} i(t)\frac{d(L(t)i(t))}{dt} dt\\
& = &  - \int_{t_0}^{t_1} L(t)i(t) \frac{d}{dt}\left( i(t) \right)dt \\ 
&=& \frac12 \int_{t_0}^{t_1} \dot L(t)i^2(t)dt.
\end{eqnarray*}
Again $\mathcal{E}$ can be negative, 
e.g.\ with $L(t)$ and $i(t)$ chosen as $k(t)$ and $x(t)$ in Example~\ref{ex:2}.
Hence Definition~\ref{def:passive} cannot be satisfied,
so the device law is active.
\end{example}

\subsection{Electrical adjustable transformer}

In analogy with the mechanical case, lossless adjustable capacitors and inductors would be realisable if it was possible to build a lossless adjustable ideal transformer.  The following governing equations can be envisaged:
\begin{subequations}\label{eq:relations}
\begin{align}\label{eq:relation_of_vs}
v_2(t) &=m(t) v_1(t) \\ \label{eq:relation_of_is}
i_2(t) &=-m^{-1}(t) i_1(t)
\end{align}
\end{subequations}
where $(v_k, i_k)$ are the voltage-current pairs for the two ports and $m(t)$ is an adjustable ratio.  We note that $v_1(t)i_1(t) + v_2(t) i_2(t) = 0$, so the device would be ``instantaneously lossless'', and indeed such a device law satisfies the generalisation of Definition~\ref{def:lossless} (losslessness) to multi-ports.  

We observe that a lossless adjustable capacitor (inductor) would result by terminating one of the ports of the transformer (\ref{eq:relation_of_vs})--(\ref{eq:relation_of_is}) with an ordinary capacitor (inductor).  For example, if we terminate the second port with a unit capacitor, which means that $i_2 = - \dot v_2$, then we find that:
\begin{equation}
i_1(t) = m(t)  \frac{d}{dt} \left( m(t)v_1(t) \right)
\end{equation}
which is analogous to the varinerter in the force-current analogy between mechanical and electrical devices.

\begin{figure}[ht]
\centering
%\begin{tikzpicture}[scale=1]  
%\tikzset{
%   ultra thin/.style= {line width=0.1pt},
%   very thin/.style=  {line width=0.2pt},
%   thin/.style=       {line width=.4pt},
%   semithick/.style=  {line width=0.6pt},
%   thick/.style=      {line width=.9pt},
%   very thick/.style= {line width=2pt},
%   ultra thick/.style={line width=0.9pt}
%}
%\draw [thick] [variable=\t, domain=0:110, samples=500, smooth] plot ({sin(\t r)}, {.03*\t - .1*cos(\t r)});
%
%% left lines
%\draw[thick] (-1.5, -.1) to [short,*-](0, -.1);
%\draw[thick] (0, 3.4) to [short,-*](-1.5,3.4);
%
%% input v0
%\node at (-1.8, 1.7) {$v_1$};
%%\draw[thin] (-1.5, 0) -- (-1.5,1.5);
%\draw[-{latex},  thin] (-1.5, 0.1) -- (-1.5,3.2);
%
%% right lines
%\draw[thick] (0, -.1) to [short,-*](1.5, -.1);
%\draw[thick] (1, 2.3) to [short,-*](1.5, 2.3);
%
%% output v1
%\node at (1.8, 1.2) {$v_2$};
%\draw[-{latex},  thin] (1.5, 0.1) -- (1.5, 2.1);
%%\draw[thin] (1.5, 0) -- (1.5,1);
%
%% label A
%\node at (1.15, 2.5) {$A$};
%
%% currents
%%\draw[thin] (-2, 3.4) -- (-1.6, 3.4);
%%\draw[thin] (1.6, 2.3) -- (1.9,2.3);
%
%%current arrow and labeling
%\node at (-2, 3.7) {$i_1$};
%\node at (-1.8, .2) {$i_1$};
%\draw[-{latex},  thick] (-2.2, 3.4) -- (-1.6, 3.4);
%\draw[-{latex},  thick] (-1.65, -.1) -- (-2.2, -.1);
%
%\node at (2.1, 2.6) {$i_2$};
%\node at (1.9, .2) {$i_2$};
%\draw[-{latex},  thick] (2.2,2.3) -- (1.6, 2.3);
%\draw[-{latex},  thick] (1.65, -.1) -- (2.2,-.1);
%
%\end{tikzpicture}
\centering
\includegraphics[totalheight=4cm]{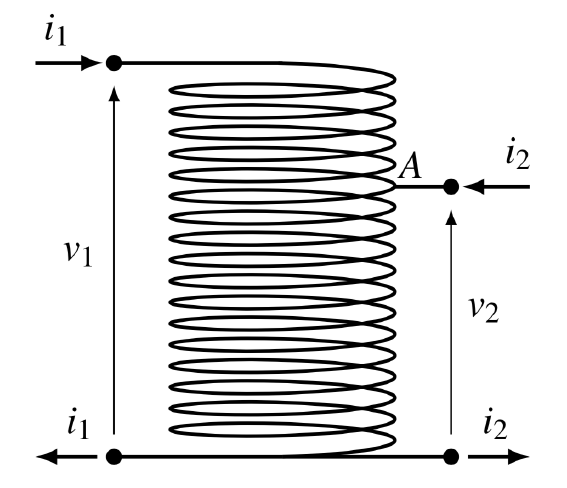}
\caption{Electrical autotransformer.}
\label{eq:electrical}
\end{figure}

Accordingly we now discuss the possibility of realising such an ideal adjustable transformer (\ref{eq:relation_of_vs})--(\ref{eq:relation_of_is}).  We note that the (ordinary) ideal electrical transformer is derived as a limit of a pair of coupled coils with perfect coupling as the inductance becomes very large.  This is a logical place to begin.  
The law specifying the electrical response of a pair of coupled coils is
\begin{eqnarray}\label{eq:coils}
\left(\begin{array}{c}v_1\\v_2\end{array}\right) &=& \frac{d}{dt}\left(
\left(\begin{array}{cc}L_1 & M\\M  & L_2\end{array}\right)
\left(\begin{array}{c}i_1\\i_2\end{array}\right)\right).
\end{eqnarray}
where $L_k$ are the self-inductances and $M$ is the mutual inductance.  In the case of perfect coupling (i.e., coils where all magnetic field lines engage both coils) we have $L_1L_2 - M^2 = 0$ and, therefore,
the inductance matrix is singular. We will consider this ideal case where we set
\begin{equation}\label{eq:inductance_matrix}
\left(\begin{array}{cc}L_1 & M\\M  & L_2\end{array}\right)=:
\left(\begin{array}{cc}L & m(t)L\\m(t)L  & m(t)^2L\end{array}\right),
\end{equation}
namely the self inductance of the first coil $L_1=L$ is assumed to be constant, the mutual inductance is $M(t)=m(t)L$, where $m(t)$ represents a time varying adjustable coupling parameter, and the self inductance of the second coil is $L_2(t)=m(t)^2L$. The coupling parameter $m(t)$ may in principle be adjusted by changing the number of turns of one of the coils, or the number of turns corresponding to the contact point on one side of an autotransformer, as shown in Figure \ref{eq:electrical}; in an autotransformer the same (portion of a) coil is shared by two circuits.

Our first observation is that the adjustable transformer defined by (\ref{eq:coils}) and (\ref{eq:inductance_matrix}) is an active device.  To see this we consider the case of $i_1=0$, namely the first port is open and the only power transfer is through the second port.  Consider a time interval $[t_0,t_1]$ in which $m(t_0)=m(t_1)$ and $i(t_0)=i(t_1)$.  Then the energy supplied to the device
\begin{eqnarray*}
\mathcal{E}  = \int_{t_0}^{t_1} i_2(t)v_2(t)dt &=& L \int_{t_0}^{t_1} i_2(t)\frac{d\left( m^2(t)i_2(t) \right)}{dt} dt\\
& = &  - L \int_{t_0}^{t_1} m^2(t) i_2(t) \frac{d}{dt}\left( i_2(t) \right)dt \\ 
&=& \frac{L}{2} \int_{t_0}^{t_1}\frac{d\left( m^2(t) \right)}{dt} i_2^2(t)dt.
\end{eqnarray*}
Again $\mathcal{E}$ can be negative, 
e.g.\ with $m^2(t)$ and $i_2(t)$ chosen as $k(t)$ and $x(t)$ in Example~\ref{ex:2}.
Hence Definition~\ref{def:passive} (passivity) cannot be satisfied.

We now explore the limiting situation in which we let $\epsilon = 1/L$ tend to zero.
%In this case, the rate of energy stored in the magnetic field is
%\[
%v_1i_1+v_2i_2=\frac12 \frac{d}{dt} L(i_1+m i_2)^2.
%\]
From (\ref{eq:coils}), (\ref{eq:inductance_matrix}),
\begin{eqnarray}
\frac{1}{L}\left(\begin{array}{c}v_1\\v_2\end{array}\right)
%&=& \frac{d}{dt}\left(
%\left(\begin{array}{c}1\\ m\end{array}\right)
%\left(\begin{array}{cc}1,& m\end{array}\right)
%\left(\begin{array}{c}i_1\\i_2\end{array}\right)\right)\\
&=& 
\frac{d}{dt}\left(\begin{array}{c}i_1+mi_2\\
m(i_1+mi_2)\end{array}\right)  \label{eq:ddelta}
\end{eqnarray}
from which we deduce that
\begin{subequations}\label{eq:relations1}
\begin{align}\label{eq:relation_of_vs1}
v_2-mv_1&=L\dot m \gamma , \\ \label{eq:relation_of_is1}
\dot{\gamma}&= v_1/L
\end{align}
\end{subequations}
where $\gamma:=i_1+mi_2$.  It is evidently not straightforward to deduce that $\gamma = 0 $ (i.e.\ that (\ref{eq:relation_of_is}) holds) from (\ref{eq:relation_of_is1}).  Even if $\gamma (t_0) = 0$, since $\dot\gamma(t) = o(\epsilon)$, there could be slow drift in $\gamma$.  This could mean that the right hand side of (\ref{eq:relation_of_vs1}) is non-negligible which would prevent (\ref{eq:relation_of_vs}) from holding when $\dot m$ is non-zero.

The above considerations show that the physical implementation of a lossless adjustable electrical transformer to realise the laws (\ref{eq:relation_of_vs})--(\ref{eq:relation_of_is}) is not a simple matter. Industrial implementations of variable transformers, such as the variac, where the contact point $A$ slides vertically, effectively shorts loops as the contact point is being repositioned to correspond to different coupling ratios. An alternative option to move the contact point $A$ displayed in Figure \ref{eq:electrical} so as to slide along the coil (as in a balustrade) does not work either. In such a scheme, a wire with the contact point $A$ as its tip would extend inside the coil as it slides through the opposite side of turns. Nonzero magnetic field lines will then exert forces that need to be overcome requiring work to be done.  %We leave it as an open question whether a lossless adjustable electrical transformer is capable of physical realisation.

In the next section we will indicate an alternative approach to construct a lossless adjustable transformer as envisaged in (\ref{eq:relation_of_vs})--(\ref{eq:relation_of_is}), though we will begin first by considering the possibility to construct lossless adjustable inductors and capacitors without resort to such a transformer. 

\subsection{Canonical device laws: electrical elements}

Before considering the matter of physical realisability we formally define device laws as follows:
\begin{eqnarray}
v&= \ell \frac{d}{dt} (\ell i) &\hspace*{10pt}\mbox{(varinductor)}\\
i &= c \frac{d}{dt} (c v) &\hspace*{10pt}\mbox{(varcapacitor)}
\end{eqnarray}
where $(v,i)$ are the terminal voltage and current of an electrical one-port and $\ell (t)$, $c(t)$ are adjustable parameters.  An internal energy may be defined by $\frac{1}{2} \ell^2 i^2$ and $\frac{1}{2} c^2v^2$ respectively, which shows that the device laws are lossless according to Definition~\ref{def:lossless}.  

One approach to the construction of varinductors and varcapacitors is to make use of a mechanical-electrical transducer to convert the mechanical rotary varspring or varinerter into electrical devices.   Consider an ideal DC permanent magnet motor-generator with
\begin{eqnarray*}
v & = & k_E \omega ,\\
T & = & k_T i,
\end{eqnarray*}
where $k_E$ and $k_T$ are the voltage and torque constants satisfying $k_E = k_T$ in SI units.  If this is connected across the terminals of a rotary varsping or varinerter then a varinductor or varcapacitor respectively is obtained.  We take this as a justification that it is possible to physically realise the varinductor and varcapacitor without resort to an internal power source.  We leave it as an open question whether more direct methods are possible.

Finally in this section we return to the question of the physical realisability of the adjustable electrical transformer.  We simply point out that if we connect an ideal DC permanent magnet motor-generator to both shafts of the mechanical adjustable transformer (\ref{eq:adjtran1})--(\ref{eq:adjtran2}) (see Fig.~\ref{fig:balls}) then we obtain a realisation of the adjustable electrical transformer (\ref{eq:relation_of_vs})--(\ref{eq:relation_of_is}) without an internal power source.  Again we leave it as an open question whether there is a more direct physical realisation.

%\section{Energy storage functions}

\section{Conclusion}

We have shown that none of the commonly assumed device laws for adjustable springs or inerters or variants are lossless, and indeed all are non-passive, i.e.\ active.   Using an idealised mechanical arrangement of a lever with moveable fulcrum device laws were derived for lossless adjustable springs and inerters.  A physical implementation of the moveable fulcrum concept without internal power source was presented for the canonical lossless adjustable spring and inerter which were named the varspring and varinerter.  A method for physical implementation of rotary varsprings and varineters was presented.  The paper included a discussion of the analogous device laws in the electrical domain.

%\section*{References}
%
\bibliography{mybibfile}
\bibliographystyle{IEEEtran}

%\newpage

\end{document}